\documentclass[10pt,twocolumn, nofootinbib]{revtex4-2}

\usepackage{amsmath}
\usepackage{amsfonts}
\usepackage{graphicx}
\usepackage{enumitem}
\usepackage{hyperref}
\hypersetup{
	colorlinks=true,
	citecolor=blue,
	urlcolor=blue,
	linkcolor=blue
}
\begin{document}

\title{Classical mechanics and infinitesimal reducibility }
\author{Gabriele Carcassi, Christine A. Aidala}
\affiliation{Physics Department, University of Michigan, Ann Arbor, MI 48109}

\date{\today}

\begin{abstract}
We briefly show how classical mechanics can be rederived and better understood as a consequence of three assumptions: infinitesimal reducibility, deterministic and reversible evolution, and kinematic equivalence.
\end{abstract}

\maketitle

\section{Introduction}

The overall argument (see \cite{AoPPhy1,aop-book} for more details) can be summed up in the following diagram that can be used as a guide throughout this note.

\begin{figure}[h]
	\includegraphics[width=\columnwidth]{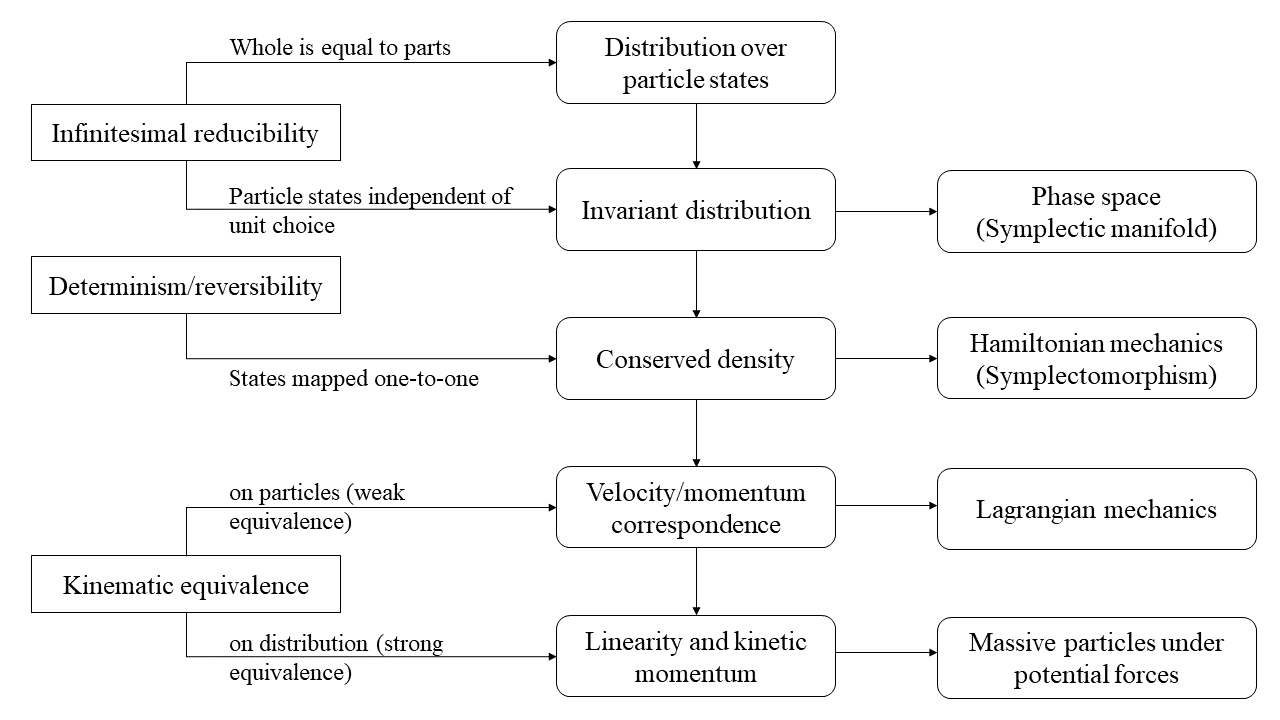}
\end{figure}

The three assumptions lie on the left column. Each assumption leads to one or two key insights that progressively lead to the physical concepts in the middle column. Each of these is then mapped to its corresponding formal framework on the right.

\section{Infinitesimal reducibility}

\textbf{Infinitesimal reducibility assumption:} \emph{the state of the system is reducible to the state of its infinitesimal parts.} That is, giving the state of the whole system is equivalent to giving the state of its parts, which in turn is equivalent to giving the state of its subparts and so on.

Under this assumption, the state of the whole is a distribution over the states of the parts. More precisely, let $\mathcal{C}$ be the state space for the whole system. We call particle the limit of recursive subdivision. Let $\mathcal{S}$ be the state space for the particles. Then for each $c \in \mathcal{C}$ we can find one and only one $\rho_c : \mathcal{S} \to \mathbb{R} $ that describes the state of its parts. That is, the state of the whole is a distribution over the infinitesimal parts. If $\mathcal{S}$ is a manifold, $\int_U \rho_c d\mathcal{S}$ gives us the fraction of the system whose parts are in the region $U$ of the state space.

The next step takes the rest of the section. We need to show that $\mathcal{S}$ has the structure of phase space, with its conjugate variables. Mathematically, $\mathcal{S}$ is a symplectic manifold. The key insight is that, on a manifold, $\rho_c$ must transform both as a scalar function (the value must depend on the point and not on the coordinates) and as a density. Classical phase space is the only space that allows these invariant distributions.

The key concept is to keep track of the unit system, so we need precise terminology. We call state variables a set of quantities $\xi^a$ that fully identify a state.\footnote{In mathematical terminology, these are the coordinates of the manifold.} That is, we can write each state $s(\xi^a)$ as a function of the state variables. We call coordinate $q \in \xi^a$ a particular variable that defines a unit. The key problem is to understand how many coordinates we can have for a given set of state variables.

We start with the simplest case, where one coordinate is sufficient to define the unit system, which means the following four conditions:
\begin{enumerate}[noitemsep]
	\item the state variables can be written as $\xi^a = \{ q, k^i \}$
	\item we can arbitrarily change coordinate to $\hat{q}=\hat{q}(q)$
	\item a change of coordinate induces a unique change over the remaining state variables $\hat{k}^j = \hat{k}^j(q, k^i)$
	\item the density is the same regardless of the coordinates used.
\end{enumerate}

Now we show that there can only be one $k^i$. Suppose we change unit $\hat{q}=\hat{q}(q)$. Call the new units $\hat{\xi}^b = \{ \hat{q}, \hat{k}^j\}$. We have $\rho_c(\hat{\xi}^b)=\rho_c(s(\hat{\xi}^b))=\rho_c(s(\xi^a))=\rho_c(\xi^a) = \left|\frac{\partial \hat{\xi}^b}{\partial \xi^a} \right| \rho_c(\hat\xi^b)$. Therefore the Jacobian $\left|\frac{\partial \hat{\xi}^b}{\partial \xi^a} \right|$ must be equal to 1. Note that, since $\hat{q}$ depends only on $q$, $\left|\frac{\partial \hat{\xi}^b}{\partial \xi^a} \right| = \left|\frac{\partial \hat q}{\partial q} \right|\left|\frac{\partial \hat{k}^j}{\partial k^i} \right|$. Suppose there is only one variable. Then we would have $\left|\frac{\partial \hat q}{\partial q} \right| = 1$. But this would mean the unit change cannot be arbitrary. Therefore we must have two or more variables and therefore $\left|\frac{\partial \hat{k}^b}{\partial k^a} \right| = \left|\frac{\partial q}{\partial \hat q} \right|$. This puts a constraint only on the determinant of the transformation. Suppose there are three or more variables. This constraint is not enough to recover the transformation uniquely and therefore $\hat q$  would not fully define the units for all other state variables. This means there must be two variables: $q$ and a single $k$. Coordinate independent areas and densities can only be defined on a two-dimensional manifold.

Now we generalize to multiple coordinates. We say two coordinates are independent if changing the units for one does not induce a change of units for the other. Now suppose our particle state space $\mathcal{S}$ is such that its units are fully defined by $n$ independent coordinates $q^i$. Suppose you change the first coordinate $q^1$ without changing the others. Then we will find a variable $k_1$ that changes as before so that the densities are invariant. Now change the second coordinate $q^2$ in the same way while also fixing $k_1$. Then we find a corresponding $k_2$. We can proceed in the same way until we exhaust all coordinates, which must also mean that there are no state variables left. We find that $\mathcal{S}$ is $2n$-dimensional, and the state variables are $\xi^a = \{ q^i, k_i \}$. We define an independent degree of freedom as the space charted by a pair of such variables.

We can dress the result a bit more formally, and show that $\mathcal{S}$ is a symplectic manifold. To characterize marginal distributions on each degree of freedom, we want to define integrals of the form $\int_{\Sigma} \rho_c \omega(d\Sigma)$ such that the density $\rho_c$ is invariant. Therefore we need a two-form $\omega$ that assigns an infinitesimal area to each infinitesimal surface, and we need $\omega$ to be invariant. Because degrees of freedom are, at least locally, independent, the total number of states in a volume is the product of the possible configurations of each degree of freedom. This means the volume form is proportional to $\omega^n$. This cannot be degenerate (i.e.~it must be nonzero for each infinitesimal volume) since all regions of phase space must, by definition, contain some states. Therefore $\omega$ itself cannot be degenerate. As the degrees of freedom are independent, the number of states on a surface does not change if we translate it across independent degrees of freedom. If we imagine a parallelepiped, the integral over the surface must be zero (integrals over opposite sides are equal and opposite). Therefore $\mathcal{S}$ must come equipped with a two-form $\omega$ that is closed and not degenerate and therefore $\mathcal{S}$ is a symplectic manifold. By convention, we set $\omega = \hbar dq^i \wedge dk_i = dq^i \wedge dp_i$ where $p_i = \hbar k_i$.

Phase space (i.e.~a symplectic manifold) is the only type of manifold that is able to support coordinate invariant distributions, which are required to describe an infinitesimally reducible system.

\section{Deterministic and reversible evolution}

\textbf{Deterministic and reversible evolution assumption:} \emph{given the present state of the system, all future (determinism) and past (reversibility) states are uniquely identified.}

We first apply the assumptions to the motion of a single particle. Let $\lambda : \mathbb{R} \to \mathcal{S}$ be the evolution over time of the state of a particle. Under the assumption, this will be uniquely identified by the initial state $s_0 = \lambda(t_0)$ at the initial time $t_0$. Secondly, we apply the assumption to the density in the sense that all the particles that start with the same initial state must end up in the same final state and vice-versa. That is, if $\rho(\lambda(t_0), t_0)$ is the density associated to the initial particle at the initial time, we must have that $\rho(\lambda(t), t) = \rho(\lambda(t_0), t_0)$: the density must remain the same throughout the evolution.

Now consider the integral $\int_{\Sigma} \rho \omega(d\Sigma)$. Both the region and the density will be mapped in time to $\hat{\Sigma}$ and $\hat{\rho}$ respectively. The fraction of the system found in the new region will have to be the same as the one found in the old region. That is, $\int_{\Sigma} \rho \omega(d\Sigma) = \int_{\hat{\Sigma}} \rho \hat{\omega}(d\hat{\Sigma})$. Since both the integral and the density have to remain constant during the evolution, then $\omega$ will need to remain the same. That is, the areas in phase space must be mapped to areas of equal size and independent degrees of freedom must be mapped to independent degrees of freedom (or volumes would not be mapped to equal volumes). The evolution is a symplectomorphism and corresponds to Hamiltonian evolution. Intuitively, this is the inverse of Liouville's theorem: instead of positing Hamiltonian evolution and finding conservation of areas and volumes, deterministic and reversible evolution imposes the conservation of areas and volumes which leads to Hamiltonian evolution.

The argument can also be constructed through statistical concepts (i.e.~determinism and reversibility means conservation of variance), thermodynamic concepts (i.e.~determinism and reversibility means only the state of the system is important for the evolution; the system is therefore isolated and must conserve energy) or information theoretic consideration (i.e.~under deterministic and reversible evolution the amount of information does not change, so information entropy has to be conserved).

\section{Kinematic equivalence}

\textbf{Kinematic equivalence assumption:} \emph{the motion of the system (i.e.~trajectories in physical space-time) is enough to recover its dynamics (i.e.~evolutions in state space) and vice-versa.}

First, as before, we apply the assumption to the particles, which means that for every evolution in phase space there should be one and only one trajectory. Note that each space variable $x^i$ is a coordinate, i.e. a state variable that defines a unit. In fact, the trajectories can be fully described by those units and only those units. So we can say that $q^i=x^i$, each $q^i$ will be paired with a conjugate $p_i$ and each state $\{q^i, p_i\}$ will be mapped to one and only one trajectory. At each point $x^i$, then, infinitely many trajectories must pass, one for each combination of $\{p_i\}$. Since the trajectories are differentiable in $x^i$, we can define a velocity $v^i = d_t x^i$. If the equations of motion were such that $v^i=v^i(q^i)$, then kinematic equivalence would fail as the full trajectory would be determined only by $q^i$. So we must have $v^i=v^i(q^i, p_i)$. The relationship must be invertible or kinematic equivalence would fail. At any given time, then, we must have the following relationship:
\begin{equation}
\begin{aligned}
x^i &= q^i \\
v^j &= d_t x^j = v^j(q^i, p_k)
\end{aligned}
\end{equation}

Let us call weak equivalence the notion that $v^j(q^i, p_k)$ must be invertible at every $q^i$ and therefore we can write $p_k= p_k(x^i, v^j)$ as a function of position and velocity. In this case, the Jacobian matrix  $\frac{\partial v^j}{\partial p_k}$ must be invertible. Since $v^j = d_t q^j = \frac{\partial H}{\partial p_j}$, the Hessian $\frac{\partial^2 H}{\partial p_k \partial p_j}$ must be nonzero everywhere, and therefore must have the same sign which we take to be positive. In this case, and only in this case, we can construct a Lagrangian from a Hamiltonian:
\begin{equation}
	L(x^i, v^j) = v^k p_k (x^i, v^j) - H(q^i(x^i), p_k(x^i, v^j))
\end{equation}
These are also exactly the cases where the Lagrangian, using the principle of minimal action, leads to a unique solution.

Now we look at the whole distribution and how it can be expressed as a function of position and velocity. We have $\rho(q^i, p_j) = |J| \rho(x^i, v^j) = \left|\frac{\partial v^i}{\partial p_j}\right| \rho(x^i, v^j)$ since
\begin{equation}
\begin{aligned}
|J| &= \begin{vmatrix}
\frac{\partial x^i}{\partial q^j} & \frac{\partial x^i}{\partial p_j} \\
\frac{\partial v^i}{\partial q^j} & \frac{\partial v^i}{\partial p_j}
\end{vmatrix}
= \begin{vmatrix}
\delta^i_j & 0 \\
\frac{\partial v^i}{\partial q^j} & \frac{\partial v^i}{\partial p_j}
\end{vmatrix} \\
&= \left|\delta^i_j\right| \left|\frac{\partial v^i}{\partial p_j}\right| - \left|0\right| \left|\frac{\partial v^i}{\partial q^j}\right|
= \left|\frac{\partial v^i}{\partial p_j}\right|.
\end{aligned}
\end{equation}
Note that while the value given by $\rho(q^i, p_j)$ is coordinate independent, the value given by $\rho(x^i, v^j)$ depends on the choice of coordinate through $\left|\frac{\partial v^i}{\partial p_j}\right|$. If $x^i$ truly sets the unit system by itself, then $\left|\frac{\partial v^i}{\partial p_j}\right|$ must be a function of position only. Similar considerations will also hold for marginal distributions (i.e.~distributions on a subset of the coordinates) which means all components of $\frac{\partial v^i}{\partial p_j}$ must be a function of position only. We set:
\begin{equation}
\begin{aligned}
\frac{\partial v^i}{\partial p_j} = \frac{1}{m} g^{ij} \\
\frac{\partial p_j}{\partial v^i} = m g_{ji}
\end{aligned}
\end{equation}
where $m$ is the unit conversion constant between velocity and conjugate momentum while $g_{ij}$ represents the linear dependency.

If we integrate, we have:
\begin{equation}
\begin{aligned}
v^i = \frac{1}{m} g^{ij}(p_j - A_j) \\
p_j = m g_{ji} v^i + A_j
\end{aligned}
\end{equation}
where $A_j$ are arbitrary functions. Note that:
\begin{equation}
v^i = d_t q^i = \frac{\partial H }{\partial p_i} = \frac{1}{m} g^{ij}(p_j - A_j) \\
\end{equation}
We integrate yet again and find:
\begin{equation}
H = \frac{1}{2m} (p_j - A_j) g^{ij}(p_j - A_j) + V \\
\end{equation}
where V is another arbitrary function. We recognize this as the Hamiltonian for massive particles under potential forces.

\bibliography{bibliography}

\begin{thebibliography}{2}%
\makeatletter
\providecommand \@ifxundefined [1]{%
 \@ifx{#1\undefined}
}%
\providecommand \@ifnum [1]{%
 \ifnum #1\expandafter \@firstoftwo
 \else \expandafter \@secondoftwo
 \fi
}%
\providecommand \@ifx [1]{%
 \ifx #1\expandafter \@firstoftwo
 \else \expandafter \@secondoftwo
 \fi
}%
\providecommand \natexlab [1]{#1}%
\providecommand \enquote  [1]{``#1''}%
\providecommand \bibnamefont  [1]{#1}%
\providecommand \bibfnamefont [1]{#1}%
\providecommand \citenamefont [1]{#1}%
\providecommand \href@noop [0]{\@secondoftwo}%
\providecommand \href [0]{\begingroup \@sanitize@url \@href}%
\providecommand \@href[1]{\@@startlink{#1}\@@href}%
\providecommand \@@href[1]{\endgroup#1\@@endlink}%
\providecommand \@sanitize@url [0]{\catcode `\\12\catcode `\$12\catcode
  `\&12\catcode `\#12\catcode `\^12\catcode `\_12\catcode `\%12\relax}%
\providecommand \@@startlink[1]{}%
\providecommand \@@endlink[0]{}%
\providecommand \url  [0]{\begingroup\@sanitize@url \@url }%
\providecommand \@url [1]{\endgroup\@href {#1}{\urlprefix }}%
\providecommand \urlprefix  [0]{URL }%
\providecommand \Eprint [0]{\href }%
\providecommand \doibase [0]{https://doi.org/}%
\providecommand \selectlanguage [0]{\@gobble}%
\providecommand \bibinfo  [0]{\@secondoftwo}%
\providecommand \bibfield  [0]{\@secondoftwo}%
\providecommand \translation [1]{[#1]}%
\providecommand \BibitemOpen [0]{}%
\providecommand \bibitemStop [0]{}%
\providecommand \bibitemNoStop [0]{.\EOS\space}%
\providecommand \EOS [0]{\spacefactor3000\relax}%
\providecommand \BibitemShut  [1]{\csname bibitem#1\endcsname}%
\let\auto@bib@innerbib\@empty
\bibitem [{\citenamefont {Carcassi}\ \emph {et~al.}(2018)\citenamefont
  {Carcassi}, \citenamefont {Aidala}, \citenamefont {Baker},\ and\
  \citenamefont {Bieri}}]{AoPPhy1}%
  \BibitemOpen
  \bibfield  {author} {\bibinfo {author} {\bibfnamefont {G.}~\bibnamefont
  {Carcassi}}, \bibinfo {author} {\bibfnamefont {C.~A.}\ \bibnamefont
  {Aidala}}, \bibinfo {author} {\bibfnamefont {D.~J.}\ \bibnamefont {Baker}},\
  and\ \bibinfo {author} {\bibfnamefont {L.}~\bibnamefont {Bieri}},\ }\bibfield
   {title} {\bibinfo {title} {From physical assumptions to classical and
  quantum {H}amiltonian and {L}agrangian particle mechanics},\ }\href
  {https://doi.org/10.1088/2399-6528/aaba25} {\bibfield  {journal} {\bibinfo
  {journal} {Journal of Physics Communications}\ }\textbf {\bibinfo {volume}
  {2}},\ \bibinfo {pages} {045026} (\bibinfo {year} {2018})}\BibitemShut
  {NoStop}%
\bibitem [{\citenamefont {Carcassi}\ and\ \citenamefont
  {Aidala}(2021)}]{aop-book}%
  \BibitemOpen
  \bibfield  {author} {\bibinfo {author} {\bibfnamefont {G.}~\bibnamefont
  {Carcassi}}\ and\ \bibinfo {author} {\bibfnamefont {C.~A.}\ \bibnamefont
  {Aidala}},\ }\href {https://doi.org/10.3998/mpub.12204707} {\emph {\bibinfo
  {title} {Assumptions of Physics}}}\ (\bibinfo  {publisher} {Michigan
  Publishing},\ \bibinfo {year} {2021})\BibitemShut {NoStop}%
\end{thebibliography}%

\end{document}